\documentclass[10pt,a4paper,oneside,onecolumn]{article}

\usepackage{ICROMA}
\usepackage{enumerate}

\usepackage{amsmath,amssymb}
\usepackage{graphicx}
\usepackage{hyperref}
\usepackage{caption}
\usepackage{subcaption}
\usepackage{algorithm}
\usepackage{algpseudocode}
\usepackage{tikz}
\usepackage{easyReview}
\usetikzlibrary{shapes.geometric, arrows, calc}
\usepackage{natbib}
\usepackage{multirow}

\title{Scalable, Technology-Agnostic Diagnosis and Predictive Maintenance for Point Machine using Deep Learning.}
\author{
        Ci Ruixiang\textsuperscript{2,*}, 
        Di Santi Eduardo\textsuperscript{1,*}, 
        Lefebvre Clément\textsuperscript{1}, 
        Mijatovic Nenad\textsuperscript{1},\\
        Pugnaloni Michele\textsuperscript{1}, 
        Jonathan Brown\textsuperscript{1},
        Martín Victor\textsuperscript{1}, 
        Saiah Kenza\textsuperscript{1}
}
\affiliation{
         \textsuperscript{1}Digital and Integrated Systems, Alstom\\
         \textsuperscript{2}Innovation and Smart Mobility, Alstom\\
\noindent
E-mails: \href{mailto:ci.ruixiang@alstomgroup.com}{ci.ruixiang@alstomgroup.com}, 
\href{mailto:eduardo.di-santi@alstomgroup.com}{eduardo.di-santi@alstomgroup.com}, 
\href{mailto:clement.lefebvre-renard@alstomgroup.com}{clement.lefebvre-renard@alstomgroup.com}, 
\href{mailto:nenad.mijatovic@alstomgroup.com}{nenad.mijatovic@alstomgroup.com}, 
\href{mailto:michele.pugnaloni@alstomgroup.com}{michele.pugnaloni@alstomgroup.com}, 
\href{mailto:jonathan.brown@alstomgroup.com}{jonathan.brown@alstomgroup.com}, 
\href{mailto:victor-andres.martin@alstomgroup.com}{victor-andres.martin@alstomgroup.com}, 
\href{mailto:kenza.saiah@alstomgroup.com}{kenza.saiah@alstomgroup.com}
}
\begin{document}

\maketitle

\begin{abstract}
    The Point Machine (PM) is a critical piece of railway equipment that switches train routes by diverting tracks through a switchblade. As with any critical safety equipment, a failure will halt operations leading to service disruptions; therefore, pre-emptive maintenance may avoid unnecessary interruptions by detecting anomalies before they become failures. Previous work relies on several inputs and crafting custom features by segmenting the signal. This not only adds additional requirements for data collection and processing, but it is also specific to the PM technology, the installed locations and operational conditions limiting scalability.  Based on the available maintenance records, the main failure causes for PM are obstacles, friction, power source issues and misalignment. Those failures affect the energy consumption pattern of PMs, altering the usual (or healthy) shape of the power signal during the PM movement. In contrast to the current state-of-the-art, our method requires only one input. We apply a deep learning model to the power signal pattern to classify if the PM is nominal or associated with any failure type, achieving >99.99\% precision, <0.01\% false positives and negligible false negatives. Our methodology is generic and technology-agnostic, proven to be scalable on several electromechanical PM types deployed in both real-world and test bench environments. Finally, by using conformal prediction the maintainer gets a clear indication of the certainty of the system outputs, adding a confidence layer to operations and making the method compliant with the ISO-17359 standard. 
\end{abstract}


\section{Introduction}
Point Machines (PM) are critical elements in railway infrastructure. They physically modify the track by acting on a movable component called a blade, which connects the rails to an alternative track, thereby changing the direction of the train (see Figure \ref{fig:point_machine_switch}):

\begin{figure}[H]
    \centering
    \begin{tikzpicture}

        \draw[thick] (-3, 0) -- (3, 0);  
        \draw[thick] (-3, -0.3) -- (3, -0.3);  
        
        \node at (-4.5, 0) {Main Track};
        
        \draw[thick] (0, 0) -- (2.5, 2);  
        \draw[thick] (0, -0.3) -- (2.5, 1.7);  

        \node at (3.7, 2) {Alternate Route};
        
        \draw[ultra thick, dashed] (0, 0) -- (1, 0.7);  
        
        \node at (1.0, 1) {Blade};

        \draw[->, thick] (-2, 0.1) -- (-1, 0.1);  
        \node at (-1.5, 0.5) {Train Direction};

        \draw[dotted, thick] (1.9, -0.3) arc[start angle=10, end angle=90, radius=1.2];

    \end{tikzpicture}
    \caption{PM operation: The blade (dashed) moves to connect the main track to the alternate route}
    \label{fig:point_machine_switch}
\end{figure}
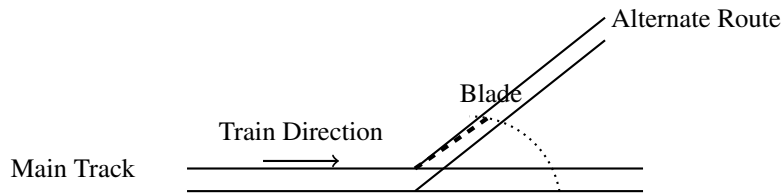
A PM failure may lead to service disruptions that quickly propagate to the networked railway system, causing downtime that impacts service satisfaction, creates costly delays and even potential penalties to the operator. For this reason, being able to detect failures before they happen and carry on preventive actions, known as predictive maintenance, is crucial to plan and execute the proper interventions to ensure safe and reliable operations.
Implementing effective predictive maintenance requires understanding that usually, failures evolve from anomalies, i.e., a condition still within operational parameters but different to the normal operation, to a condition outside those parameters, which is a failure.
Building on previous signal segmentation approaches\citep{inproceedings3}, our method focuses on isolating key segments within signal trajectories using a single primary input, then applying deep learning layers to generalise across different PM types, installations and configurations.
By observing other types of electromechanical point machines, we found that the same segmentation can be applied: 
\begin{figure}[H]
    \centering
    \includegraphics[width=1\linewidth]{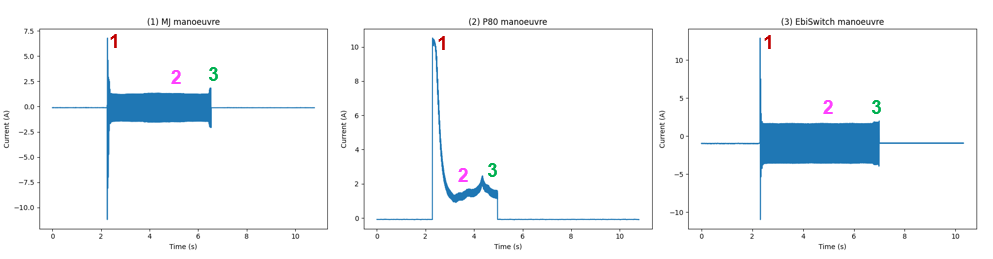}
    \caption{Typical nominal Manoeuvres of (1) MJ, (2) P80 and (3) EbiSwitch}
    \label{fig:manoeuvres}
\end{figure}
In Figure \ref{fig:manoeuvres}, three stages of the manoeuvre can be observed despite different PM technologies. There is an initial power peak when the blade starts moving, a constant energy consumption as the blade moves, and another peak when the blade reaches its final position.

Our work introduces a method tailored to the unique operational patterns inherent in PM systems, which reflects standard health and performance indicators present across different PM technologies. This new method is capable of finding distortions in the normal pattern and classify the anomaly type. Finally, we integrate conformal prediction into the system to quantify the level of certainty for each prediction. This provides operators with an indication of how confident the system is on this particular prediction. This extra layer of information solves the false positive problem by quantifying the certainty of the estimation; also solves the problem of mixed anomalies by giving a probability for every fault; which improves the trustworthiness of the system. Overall, the comprehensive integration of all these methods ensures the compliance to the ISO-17359 standard \citep{iso}.

We validated our approach in both real-world and controlled test environments, achieving >99.99\% precision, <0.01\% false positives and negligible false negatives. These results indicate strong performance across diverse conditions, and so far, ongoing real-world testing has consistently confirmed the approach’s effectiveness in operational settings.

This paper is structured as follows:
\begin{enumerate}
 \item Section 2 reviews the current state of research on PM diagnostics.
 \item Section 3 presents the datasets used.
 \item Section 4 explains our methodology in detail, including the preprocessing steps and model design.
 \item Section 5 presents our experimental results and findings
 \item Section 6 concludes the paper and suggests directions for future work.
 \end{enumerate}
\section{Related Work}
The fault diagnosis of PMs has garnered significant attention using various approaches. Traditionally, threshold-based methods have been applied by setting pre-defined thresholds on the monitored parameters including the current, voltage, force and displacement \citep{inproceedings, 745106, asada2013novel, LETOT2015958}; When a certain parameter exceeds the threshold, an alarm is raised. This method often suffers from low accuracy in anomaly detection\citep{li2019fault}. To address this limitation, researchers have explored supervised learning models for binary classification with feature extraction techniques such as statistical technique \citep{inproceedings4}, Principal Component Analysis \citep{articleb}, Discrete Wavelet Transform \citep{asada2013algorithm}, Deep Wavelet Scattering Transform \citep{inproceedings3}. However, feature extraction often requires extensive customized feature engineering, making them highly specific to each PM type. Additionally, the supervised learning method poses significant challenges due to the lack of data labelling from the field and the inherent scarcity of anomalous data. There are also efforts to utilize distance between nominal and anomalous signals using methods such as Dynamic Time Warping \citep{articlea} and Shapelets \citep{sa2017replacement}. Nonetheless, the performance in detecting nuanced anomalies is still a challenge.

In recent years, the rapid advancement of deep learning has led to increasing interest in employing deep learning models \citep{8999070, 9531564} for fault diagnosis. One major advantage is its ability to learn complex patterns within the data while minimizing the requirements for manual feature engineering.

In addition, current research for PM fault diagnosis focuses on detection of anomalies \citep{10.1093/tse/tdac036}, with limited attention given to the quantification of uncertainties from the predictions. In practice, it is highly beneficial to provide the uncertainty quantification to  provide additional confidence to the maintainers.

\section{Datasets}
Proprietary datasets on three PMs - MJ, P80 and EbiSwitch - are used in this project. For MJ and P80, data are generated from test bench in a laboratory, and labelled with the anomaly type (nominal, obstacle, friction, power supply issue or misalignment of movable parts). Additionally, to validate the performance, MJ data are also collected from a revenue-service metro system. For EbiSwitch, data are collected from field in a revenue-service rapid rail system, without any data labelling. Table \ref{tab:data collection} and Table \ref{tab:data summary} summarize the proprietary datasets used for this study.

\begin{table}[H]
    \centering
    \resizebox{\textwidth}{!}{
        \begin{tabular}{|c|c|c|c|c|} \hline  
         & Data Source &  Power Supply& Number of PM Studied &Duration of Data Collection\\ \hline  
         \multirow{2}{*}{MJ}&  Test bench&  AC& 1 &9 days\\ \cline{2-5} 
         &  Field&  AC& 28 &121 days\\ \hline  
         P80&  Test bench&  DC& 1 &8 days\\ \hline  
 EbiSwitch& Field& AC&17 &60 days\\ \hline 
        \end{tabular}
    }
    \caption{Summary of data collection of the proprietary datasets}
    \label{tab:data collection}
\end{table}

\begin{table}[H]
    \centering
    \resizebox{\textwidth}{!}{
        \begin{tabular}{|l|l|l|l|l|} \hline  
\multirow{2}{*}{} &\multicolumn{4}{|c|}{Number of Valid Manoeuvres}\\   \cline{2-5}  
&MJ test bench& P80 test bench& MJ field&EbiSwitch field\\ \hline  
nominal & 356
& 263
& \multirow{5}{*}{No labelling} &\multirow{5}{*}{No labelling} \\ \cline{1-3} 
 obstacle & 274
& 503
& &\\ \cline{1-3}   
 friction & 355
& NA
& &\\ \cline{1-3}   
power supply issue & 125
& 164
& &\\ \cline{1-3}   
misalignment of movable parts & NA
& 203
& &\\ \hline  
total & 1110
& 1133
& 58407&13119\\ \hline 
        \end{tabular}
    }
    \caption{Summary of manoeuvres of the proprietary datasets}
    \label{tab:data summary} 
\end{table}

\section{Methodology}
\begin{figure}[H]
    \centering
    \includegraphics[width=0.5\linewidth]{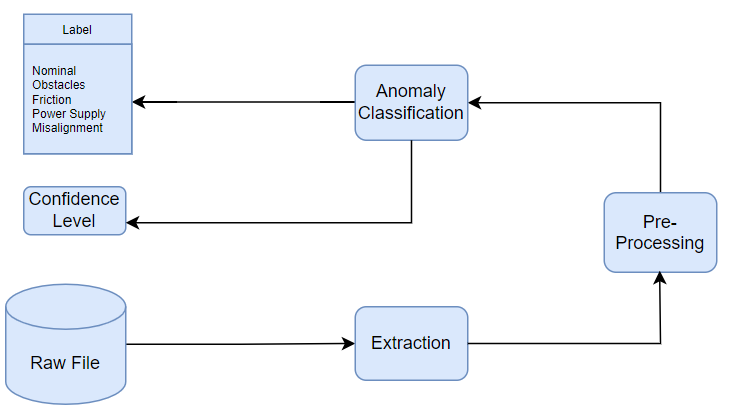}
    \caption{Methodology framework for diagnostics and predictive maintenance of PM}
    \label{fig:framework}
\end{figure}

We propose an end-to-end methodology framework for diagnostics and predictive maintenance of PM. First, the currents are extracted from the raw files. Next, a proprietary preprocessing step is applied to refine the key characteristics and reduce the dimensionality of the raw signal. This is followed by the deep learning models to classify anomalies. Lastly, conformal prediction is applied to quantify uncertainty and provide a confidence level of the predictions. Figure \ref{fig:framework} illustrates the flow of this framework.

\subsection{Preprocessing}
Data preprocessing is a critical step of feature engineering to refine the key characteristics from the raw signals of the PM manoeuvres. As raw signals may contain noise or irrelevant information, feeding them directly to the models could impact the diagnostics' performance. Moreover, data prepossessing also aids in dimensionality reduction to ensure that the neural networks can accurately learn and generalize from the underlying patterns, which leads to improved predictive capabilities. Lastly and most importantly, the preprocessing helps isolate the physical nature of the PM manoeuvres from the influence of different PM technologies and operational conditions, which is key in developing a technology-agnostic diagnostics solution.

\subsection{Deep learning for anomaly classification}
Following the preprocessing, we employ a deep learning model to classify if the signal is nominal or belongs to one of the anomaly types: obstacles, friction, power supply issues, and misalignment of movable parts. The deep learning model is trained on a dataset with labelled anomaly types. The multi-layer architecture effectively learns the complex relationships between the anomaly type and the input signal pattern. This approach aims to provide a comprehensive understanding of the nature of anomalies, facilitating the planning of maintenance activities.

\subsection{Conformal prediction for uncertainty quantification}
The trained anomaly classifier only classifies the anomaly into one of the learnt anomaly types. However, we are also interested in quantifying the uncertainties of our predictions. To address this, we employ conformal prediction to evaluate the confidence level associated with the classification outcome.

Conformal prediction is a statistical technique that provides a framework for assessing the reliability of predictions made by the models, offering calibrated prediction sets for the estimated outcomes \citep{angelopoulos2021gentle}.

Unlike standard probabilistic classifiers that output softmax probabilities, conformal prediction provides a formal guarantee: if a confidence level of 90\% is chosen, then over multiple predictions, at least 90\% of the predictions will include the correct label within the predicted set. This means that the confidence measure should not be interpreted as the probability of a single prediction being correct, but rather as the proportion of repeated predictions that will contain the true class.

The process begins with a predictive model already trained. We apply conformal prediction techniques to generate prediction intervals for new, unseen data. To achieve this, we utilise a conformal predictor that operates by assessing how well the predictive model aligns with a separate calibration dataset, which is different from the training dataset. We then calculate the conformity score based on the predefined risk level and calculation method \citep{Cordier_Flexible_and_Systematic_2023}). Based on the conformity score, we can construct a prediction set. These intervals quantify the level of uncertainty associated with each prediction, allowing stakeholders to make informed decisions based on the likelihood of various outcomes.

\textbf{Practical Interpretation for Maintenance Operations:}  
- If the conformal predictor assigns a \textbf{95\% confidence} to an "obstacle" failure type, maintainers can act with high confidence that this is the correct diagnosis.  
- If the system assigns \textbf{60\% confidence to "friction" and 40\% to "power supply"}, then the maintainer knows that both possibilities should be considered for further inspection.  

Thus, conformal prediction provides \textbf{uncertainty-aware predictions} that allow operators to prioritise interventions and assess risks associated with different diagnoses.
Conformal prediction enhances the reliability of the classification outcomes. For instance, in the event of co-occurrence of multiple anomaly types, the maintenance team will receive the confidence levels associated with a set of predictive outcomes, instead of only one predictive class. This quantification of uncertainty is crucial for making informed operational decisions in PM maintenance, as it allows prioritization of anomaly responses based on the reliability of the model's predictions.

\section{Results}

\subsection{Anomaly Classification Results}
For MJ and P80, stratified sampling with an 80\% sampling rate for each label is used to generate the training set and the remaining as the testing set. Due to the imbalanced sample size in anomaly types, we train the deep learning classifier with class weights inversely proportional to the sample size of each anomaly type. 

\begin{figure}[H]
    \centering
    \includegraphics[width=1\linewidth]{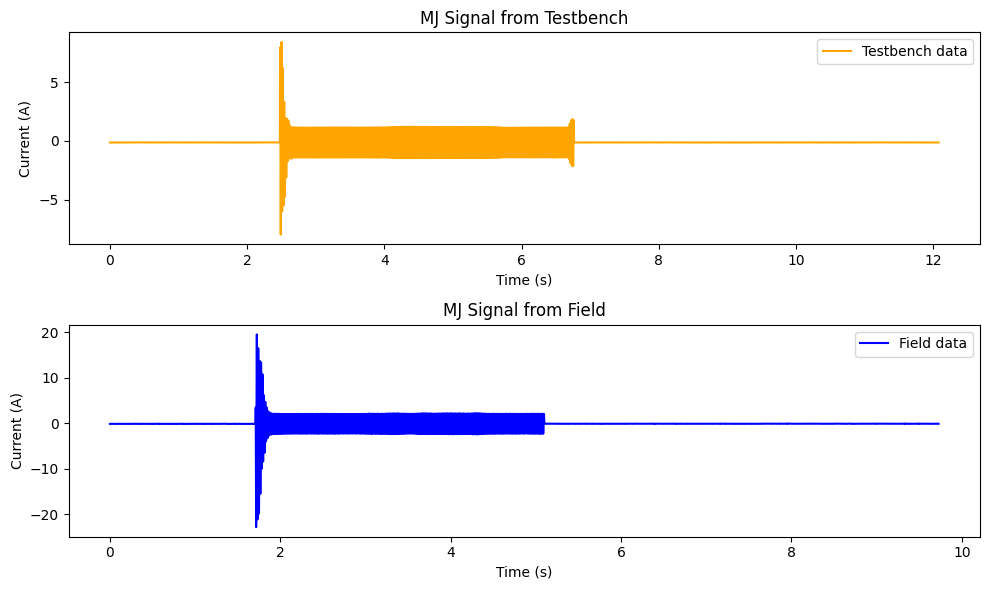}
    \caption{Field vs test bench raw signals}
    \label{fig: field vs testbench raw}
\end{figure}

\begin{figure}[H]
    \centering
    \includegraphics[width=0.7\linewidth]{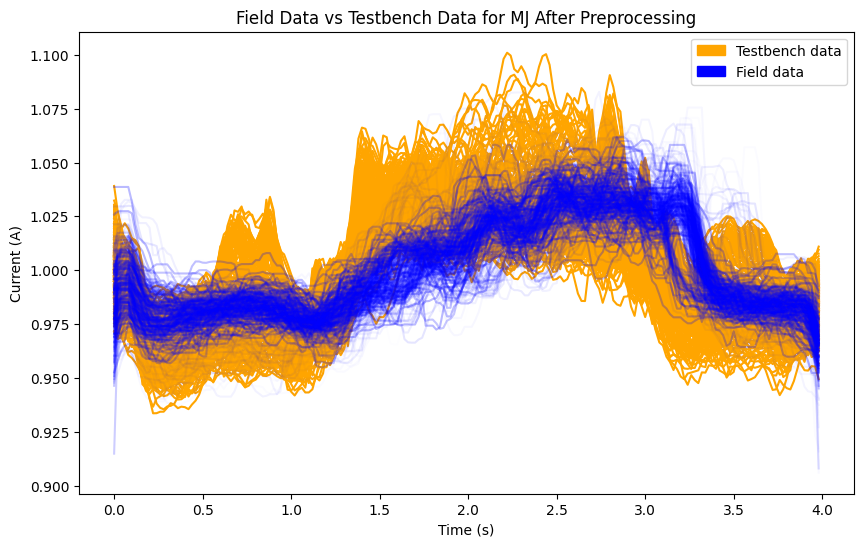}
    \caption{Field vs test bench prepossessed signals}
    \label{fig: field vs testbench}
\end{figure}

We further validate the performance of the model for MJ using real-world data collected from field point machines. Figure \ref{fig: field vs testbench raw} shows that the signals collected from the field have a similar shape as the ones from the test bench, despite having different amplitudes due to operational conditions. After applying the preprocessing method, the key characteristics of the manoeuvre are extracted and the signals look homogeneous (see figure \ref{fig: field vs testbench}).

\begin{figure} [H]
    \centering
    \includegraphics[width=0.5\linewidth]{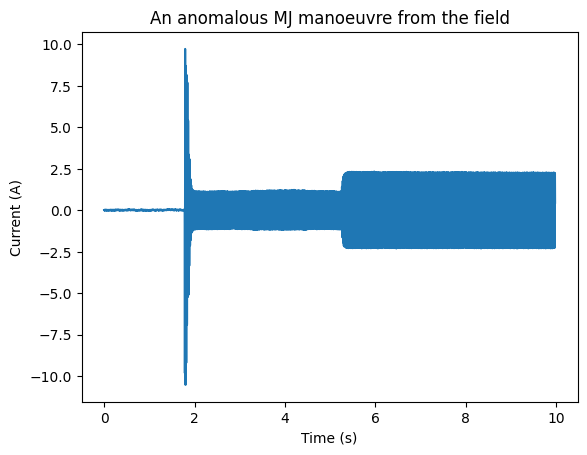}
    \caption{An anomalous MJ manoeuvre from the field}
    \label{fig:mj field anomaly}
\end{figure}

We then apply the trained anomaly detection model on the field data. There are 178 manoeuvres classified as either obstacle or friction (see figure \ref{fig:mj field anomaly} as one example). \emph{All these manoeuvres were later on confirmed and validated by the experts as genuine anomalies} .

In the case of EbiSwitch, since the data is not labelled, due to the similarity of the signal shapes between MJ and EbiSwitch, we applied the MJ classifier model to EbiSwitch for validation. 21 movements are classified as obstacles and these are validated by the expert.

We use precision, false positive rate (FPR) and false negative rate (FNR) as metrics to measure the model performance. Table 2 summarizes the performance of the anomaly classification model.

\begin{table} [h]
    \centering
    \begin{tabular}{|c|c|c|c|} \hline 
         Technology&  Precision&  FPR& FNR\\ \hline 
         MJ&  100\%&  0& 0.12\%(1)\\ \hline 
         P80&  100\%&  0& 0\\ \hline
         EbiSwitch&  100\%&  0& NA(2)\\ \hline
    \end{tabular}
    \caption{Performance of Anomaly Classification for MJ and P80}
    \label{tab:my_label}
\end{table}

\noindent
(1) Refer to testbench result.\\
(2) Revenue services, we cannot know if an anomaly exists that we are not detecting.\\

\subsection{Conformal Prediction}
So far our predictions are point estimates, i.e. they provide a single value of an estimation, rather than capturing the range of possible values or uncertainty associated with the prediction. To quantify this uncertainty, we adopt the MAPIE library \citation{Cordier_Flexible_and_Systematic_2023}for conformal prediction. We first designate half of the test set as the calibration set. Then a risk level of 0.05 and the APS method is used for calculating the conformality score and constructing the prediction set. A confidence level is thus derived based on the constructed prediction set.

\begin{figure} [H]
    \centering
    \includegraphics[width=0.7\linewidth]{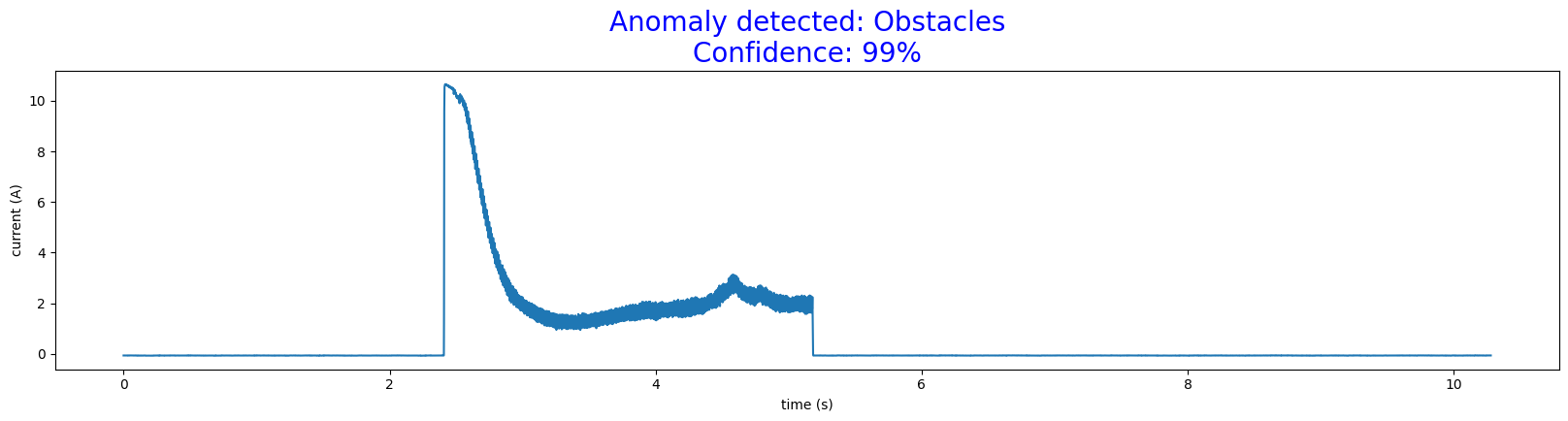}
    \caption{Example of results of the end-to-end solution with confidence level}
    \label{fig: output}
\end{figure}

Figure \ref{fig: output} illustrates an example of the results of the end-to-end solution for diagnostics and predictive maintenance, by showing the anomaly type and the confidence level, together with the raw signal of an anomalous P80 manoeuvre. This information helps improve the understanding of anomalies, helping the maintenance team make informed decisions based on the predictive result.

\section{Conclusion}
Our approach demonstrates robustness in various environments, with consistent performance for anomaly classification. Notably, the approach shows remarkable performance in real-world conditions, which validates its practical applicability within active railway systems. We build on prior work by integrating conformal prediction, which improves decision-making accuracy. Due to the preprocessing method that extracts the core phenomenon of the PM manoeuvre, our work demonstrates scalability to different PM types and environments. Future work includes extending this approach to more PM types at a larger scale, so that railway operators benefit from an efficient data-driven approach for operating and maintaining railway PMs, leading to fewer delays, lower operating costs, and improved safety.

\section*{Author Contributions}
{Ruixang Ci and Di Santi Eduardo contributed equally to this work as co-first authors.}

\bibliographystyle{unsrt}
\bibliography{references}

\end{document}